\documentclass[12pt]{article}
\usepackage{amsmath,amssymb,amsthm,amsxtra,overpic,bbm,bm,epsfig,subfigure}
\usepackage{hyperref}
\usepackage{color}
\usepackage{stackengine} 

\begin{document}
\vspace{0.2cm}

\begin{center}
{\Large\bf Modified gravity models for inflation: In conformity with observations}\\\vspace{0.2cm}
\end{center}

\vspace{0.2cm}

\begin{center}
{\bf Xinyi Zhang}~$^{a}$~\footnote{Email: zhangxinyi@ihep.ac.cn},
\quad
{\bf Che-Yu Chen}~$^{b}$~\footnote{Email: b97202056@gmail.com }
\quad
{\bf Yakefu Reyimuaji}~$^{c}$~\footnote{Email: yreyi@hotmail.com}
\\
\vspace{0.2cm}
$^a$ {\em \small Institute of High Energy Physics, Chinese Academy of Sciences, Beijing 100049, China}\\
$^b$ {\em \small Institute of Physics, Academia Sinica, Taipei 11529, Taiwan}\\
$^c$ {\em \small School of Physical Science and Technology,
Xinjiang University, Urumqi, Xinjiang 830046, China
}\\
\end{center}

\vspace{0.5cm}

\begin{abstract}
We consider a modified gravity framework for inflation by adding to the Einstein-Hilbert action a direct $f(\phi)T$ term, where $\phi$ is identified as the inflaton and $T$ is the trace of the energy-momentum tensor. The framework goes to Einstein gravity naturally when the inflaton decays out. We investigate inflation dynamics in this $f(\phi)T$ gravity (not to be confused with torsion-scalar coupled theories) on a general basis and then apply it to three well-motivated inflationary models. We find that the predictions for the spectral tilt and the tensor-to-scalar ratio are sensitive to this new $f(\phi)T$ term. This $f(\phi)T$ gravity brings chaotic and natural inflation into better agreement with data and allows a larger tensor-to-scalar ratio in the Starobinsky model.
\end{abstract}

\vspace{0.5cm}

\section{Introduction}

The Einstein general relativity (GR) theory is currently accepted as the best gravitational theory to describe our Universe. However, the standard big bang cosmological model, which is based on GR, inevitably suffers from some fundamental problems: Why is our Universe so flat (flatness problem) and how could two different spacetime patches in the Universe be causally connected after the big bang (horizon problem)? In fact, before the Universe starts to evolve following the big bang model, a period of exponentially accelerating phase in the early Universe well known as inflation~\cite{Guth:1980zm} solves the flatness and horizon problems. Furthermore, the quantum fluctuation of the inflaton field seeds the anisotropy as seen by CMB observations (for reviews, see, e.g. Refs.~\cite{Lyth:1998xn,Bassett:2005xm,Martin:2013tda}). At the present time, inflation becomes part of the standard cosmological model but the details on how our Universe inflates remains unknown. There are many competing inflationary models, among them the single-field slow-roll models are appealing as they have a minimal structure to get the job done. Regrettably, the latest Planck 2018 observations~\cite{Planck:2018jri} put some of these models in doubt. Before moving to a more complicated construction of inflationary models, we try to look at the problem from a different perspective. 

Regarding the late-time evolution of the Universe, Einstein's GR is still a well-tested theory. By including a cosmological constant, the $\Lambda$ cold dark matter model ($\Lambda$CDM) provides a good fit to observations, and, thus is the standard model of cosmology. However, it is not a complete description of the cosmological dynamics as the two dark components--dark energy and dark matter--are not fully explained. For example, the physical and/or geometrical origin of $\Lambda$ needs clarification, as well as its smallness. If one relaxes the assumption that our Universe is described by $\Lambda$CDM model, modified gravity theories which are extensions of Einstein gravity can also be used to address the aforementioned cosmological problems~\cite{Capozziello:2011et}. This can be done either by extending the geometric or the matter part or both of the GR action. There are many modified gravity theories, for example, $f(R)$ gravity~\cite{Buchdahl:1970ynr} and $f(\mathbb{T})$ teleparallel gravity~\cite{Ferraro:2006jd} where $\mathbb{T}$ is the torsion scalar. One can also consider modified theories of gravity by including nonminimal geometry-matter couplings, such as $f(R,\mathcal{L}_m)$ gravity~\cite{Nojiri:2004bi,Allemandi:2005qs,Bertolami:2007gv} and $f(R,T)$ gravity~\cite{Harko:2011kv}, where $\mathcal{L}_m$ and $T$ are the matter Lagrangian and the trace of the energy-momentum tensor, respectively. One of the important motivations of introducing these modifications is that the late-time acceleration of the Universe could be naturally explained without including any dark energy in the matter sector (for a pedagogical review, see, e.g., Ref.~\cite{Nojiri:2006ri}).

Apart from addressing the late-time acceleration, modified gravity theories are also applied to the early Universe, e.g., the inflationary era. In fact, from a phenomenological point of view, modified theories of gravity can be formulated with the motivation that they could represent some effective theories of the fundamental but unknown quantum theory of gravity under some energy cutoff. Based on this motivation, it is not surprising that these gravitational modifications could leave some imprints on the early inflationary stage. For instance,
Higgs inflation relies heavily on nonminimal couplings to the Ricci scalar. The Starobinsky model~\cite{Starobinsky:1980te} starts with adding a $R^2$ term in the action and it fits extremely well with observations when transforming to the Einstein frame. For the mentioned examples of modified gravity, the interplay with inflation has also been discussed in Refs.~\cite{Nojiri:2007as,Cognola:2007zu,Myrzakulov:2015qaa,Salvio:2017xul,Oikonomou:2017isf,Keskin:2018gev,Salvio:2019wcp,Bhattacharjee:2020jsf,Gamonal:2020itt,Salvio:2020axm} (see Ref.~\cite{Nojiri:2017ncd} for a recent review).

In this work, we propose a new form of modified gravity named $f(\phi)T$ gravity{\footnote{We emphasize once again that throughout this paper, $T$ stands for the trace of the energy-momentum tensor. Therefore, our model, i.e., the $f(\phi)T$ gravity, is completely different from the teleparallel gravity with nonminimal torsion-scalar $f(\phi)\mathbb{T}$ couplings \cite{Geng:2011aj,Otalora:2013dsa,Otalora:2013tba}.}}. It is an extension of Einstein gravity by adding a direct inflaton - energy-momentum tensor trace coupling. The action of $f(\phi)T$ gravity is
\begin{align}
S = \int \left( \displaystyle \frac{R}{2\kappa} + \alpha f(\phi) T + \mathcal{L}_m\right) \sqrt{-g}~ \mathrm{d}^4x \;,
\end{align}
where $\kappa \equiv 8\pi G \equiv 1/M_\mathrm{pl}^2$, $R$ is the Ricci scalar, $T$ is the trace of energy-momentum tensor and $\mathcal{L}_m$ is the matter Lagrangian. $f(\phi)$ is a dimensionless function of the inflaton $\phi$ satisfying $f(0)=0$ such that when inflaton decays out (its number density goes to zero), Einstein gravity is recovered. From this viewpoint, it is a modified gravity model for inflation only. This model also returns to Einstein gravity in $\alpha\rightarrow0$ limit. In this work, we consider only the simplest case: $f(\phi)=\sqrt{\kappa}\phi$.  This scenario is an extension of the simplest $f(R,T)$ modified gravity: $f(R,T) = R + 2 \kappa \alpha T$ by promoting $\alpha$ to a field and identifying it as inflaton. It is an interesting possibility of modified gravity that has not been investigated. Additionally, such a coupling can also be motivated in some semiclassical descriptions of quantum gravity frameworks \cite{Dzhunushaliev:2013nea,Liu:2016qfx,Harko:2018ayt,Chen:2021oal}.

The paper is organized as follows. In Sec.~\ref{sec:framework}, we briefly review the single-field inflationary model in Einstein gravity first and then introduce inflation dynamics in this $f(\phi)T$ gravity on a general basis. In Sec.~\ref{sec:application}, we show the effects of $f(\phi)T$ gravity on three well-motivated inflationary models: chaotic inflation, natural inflation and Starobinsky inflation. We give the analytical expressions for the slow-roll parameters as well as the observables. We show the results in comparison with the latest observation. In Sec.~\ref{sec:conclude}, we make some concluding remarks.

\section{General framework}\label{sec:framework}
\subsection{Inflation in Einstein gravity}
First, we briefly review the single-field inflationary model in the context of Einstein gravity. In GR, the action including the Einstein-Hilbert action and a matter action is
\begin{align}
S_\mathrm{GR} = \int \left( \displaystyle \frac{R}{2\kappa} + \mathcal{L}_m\right) \sqrt{-g}~ \mathrm{d}^4x \;.
\end{align}
Varying the action with respect to the metric, one finds the Einstein equation for the spacetime metric evolution 
\begin{align}
G_{\mu\nu}=R_{\mu\nu} -\displaystyle \frac{1}{2} g_{\mu \nu} R \equiv \kappa T_{\mu\nu}\;,
\end{align}
where $G_{\mu\nu}$ is the Einstein tensor and $T_{\mu\nu}$ is the energy-momentum tensor of matter present in the Universe. We have omitted the cosmological constant term that is irrelevant when discussing inflation. Consider a perfect fluid in its comoving frame, the energy-momentum tensor is
\begin{align}
T_{\mu}^\nu =\mathrm{Diag}\{-\rho,p,p,p\}\;,
\end{align}
where $\rho$ and $p$ are the energy density and pressure, respectively, and we use the metric signature $(-,+,+,+)$. Substituting the expression of the energy-momentum tensor and also the Friedmann-Robertson-Walker (FRW) metric in the Einstein equation, one can have the following two Friedmann equations for the background metric evolution as:
\begin{align}
 H^{2} &=\frac{\kappa \rho}{3} \;, \label{eq:FriedmannGR}\\
 \frac{\ddot{a}}{a} &=-\frac{\kappa}{6}\left(3 p+\rho\right)\;,\label{eq:accelerationGR}
\end{align}
where $a=a(t)$ is the scale factor, and $H\equiv\dot{a}/a$ is the Hubble function, where the dot denotes the derivative with respect to the cosmic time $t$. From the above two equations, one can get the continuity equation 
\begin{align}
\dot{\rho}+3 H\left(\rho+p\right)=0\;,\label{eq:continuity}
\end{align}
and also an expression for $\dot{H}$
\begin{align}
\dot{H} = -\frac{\kappa}{2} (\rho+p) \;.
\end{align}

Consider the simplest single-field model during inflationary era, the Universe is dominated by a scalar field $\phi$ (its potential $V(\phi)$) which contributes to the matter Lagrangian as
\begin{align}
\mathcal{L}_{m}=-\displaystyle\frac{1}{2} g^{\mu \nu} \partial_{\mu} \phi \partial_{\nu} \phi-V(\phi)=\frac{1}{2} \dot{\phi}^{2}-V(\phi)\;,
\end{align}
where in the last equality we assume that the inflaton field is spatially homogeneous, i.e., it only depends on the cosmic time $t$. The energy-momentum tensor of the inflaton field is 
\begin{align}
T_{\mu \nu}=\partial_{\mu} \phi \partial_{\nu} \phi+g_{\mu \nu}\left(\frac{1}{2} \dot{\phi}^{2}-V(\phi)\right) \;,
\end{align}
with its trace given by $T=g^{\mu \nu} T_{\mu \nu}=\dot{\phi}^{2}-4 V(\phi)$.
For a spatially homogeneous inflaton field, the energy-momentum tensor takes the form of a perfect fluid, and one can identify the energy density $\rho$ and the pressure $p$ as
\begin{align}
\rho = T_{00}=\frac{\dot{\phi}^{2}}{2}+V(\phi), \quad p g_{i j} = T_{i j}=\left(\frac{\dot{\phi}^{2}}{2}-V(\phi)\right) g_{i j}.\label{eq:rhopPhi}
\end{align}
Substituting $\rho,~p$ in Eq.~(\ref{eq:continuity}), one finds the Klein-Gordon equation for the scalar field evolution
\begin{align}
\ddot{\phi}+3H\dot{\phi}+V_{,\phi}=0\;,
\end{align}
where $V_{,\phi}\equiv dV/d\phi$. The inflation dynamics couples to the background evolution, which is governed by the following Friedmann equations:  
\begin{align}
 H^{2}&=\frac{\kappa}{3}\left(\frac{\dot{\phi}^{2}}{2}+V(\phi)\right) \;, \\
 \frac{\ddot{a}}{a}&=-\frac{\kappa}{3}\left(\dot{\phi}^{2}-V(\phi)\right) \;.
\end{align}

In the standard slow-roll inflationary model, inflation happens when the potential energy of the scalar field dominates the energy of the Universe, and it requires
\begin{align}
\frac{\dot{\phi}^{2}}{2} \ll V(\phi),\quad\ddot{\phi} \ll H\dot{\phi}\;.
\end{align}
In this slow-roll regime, the evolution equations of the scalar field and the background metric read
\begin{align}
&3H\dot{\phi}+V_{,\phi} \simeq0\;,\\
& H^{2}\simeq\frac{\kappa}{3}V(\phi) \;.
\end{align}
Differentiating these two equations and requiring that the higher-order derivatives are small, one can introduce the slow-roll parameters as
\begin{align}
\epsilon_\mathrm{E} = \displaystyle \frac{1}{2\kappa} \left(\frac{V_{,\phi}}{V} \right)^2\;, \quad
\eta_\mathrm{E} = \displaystyle \frac{1}{\kappa}\frac{V_{,\phi\phi}}{V}\;,\label{eq:etaGR}
\end{align}
where we use the subscript ``E" to mark that it is the case in Einstein gravity.
Slow-roll ends whenever $\epsilon_\mathrm{E}$ or $|\eta_\mathrm{E}|$ approaches $1$. Therefore, by counting from the end of slow-roll backwards and requiring that at least $50$ e-folds are generated, one can get the field value at the horizon crossing. With the field value at the horizon crossing and the associated slow-roll parameters, the spectral tilt and the tensor-to-scalar ratio can be expressed as follows
\begin{align}
n_s =1-6\epsilon_\mathrm{E}+2\eta_\mathrm{E}\;,\quad
r =16\epsilon_\mathrm{E}\;,\label{eq:nsrGR}
\end{align}
which can be used to compare with observations.

\subsection{Inflation in the $f(\phi)T$ gravity framework}
In our modified gravity model, we introduce an extra inflaton-energy-momentum-tensor coupling into the GR action
\begin{align}
S = \int \left( \displaystyle \frac{R}{2\kappa} + \alpha \sqrt{\kappa} \phi T + \mathcal{L}_m\right) \sqrt{-g}~ \mathrm{d}^4x \;.\label{ourmodelaction}
\end{align}
In $\alpha\rightarrow0$ limit, our model recovers Einstein gravity. We will see in the next section that, in most cases, we only need a very small value of $\alpha$ to accommodate the inflationary model predictions with observations. It is as well interesting to point out that after inflation, when inflaton decays out and its energy stops dominating the Universe, our model returns to Einstein gravity, too. As a result, there should be no worries about underlying contradictions or unwanted modifications compared to Einstein gravity after inflation. 

Varying the action in Eq.\eqref{ourmodelaction} with respect to the metric, one gets the modified Einstein equation as
\begin{align}
G_{\mu\nu}=R_{\mu\nu} -\displaystyle \frac{1}{2} g_{\mu \nu} R \equiv \kappa T_{\mu\nu}^\mathrm{(eff)}\;,\label{eq:EinsteinMG}
\end{align}
where we have defined an effective energy-momentum tensor as
\begin{align}
T_{\mu\nu}^\mathrm{(eff)} \equiv T_{\mu\nu} - 2\alpha \sqrt{\kappa} \phi \left( T_{\mu\nu} -\frac{1}{2} T g_{\mu\nu} + \Theta_{\mu\nu} \right) \;,
\end{align}
where $\Theta_{\mu\nu}$ is 
\begin{align}
\Theta_{\mu \nu} \equiv g^{\alpha \beta} \displaystyle \frac{\delta T_{\alpha \beta}}{\delta g^{\mu \nu}}=-2 T_{\mu \nu}+g_{\mu \nu} \mathcal{L}_{m}-2 g^{\alpha \beta} \frac{\delta^{2} \mathcal{L}_{m}}{\delta g^{\mu \nu} \delta g^{\alpha \beta}}\;.
\end{align}
The effective energy-momentum tensor encodes both the standard matter contributions and the modifications coming from the $\phi T$ coupling. In case of the inflaton field, $\Theta_{\mu\nu}$ reads
\begin{align}
\Theta_{\mu \nu}&= -2 T_{\mu \nu}+g_{\mu \nu} \mathcal{L}_{m}=-2 \partial_{\mu} \phi \partial_{\nu} \phi-g_{\mu \nu}\left(\frac{1}{2} \dot{\phi}^{2}-V(\phi)\right) =  - \partial_{\mu} \phi \partial_{\nu} \phi -T_{\mu\nu} \;,
\end{align}
with components $\Theta_{00}=-T_{00}-\dot{\phi}^{2}, \quad \Theta_{i j}=-T_{i j}$ and trace $\Theta = g^{\mu \nu} \Theta_{\mu \nu}=4 V(\phi)$.
From $T_{\mu\nu}^\mathrm{(eff)}$, one can further introduce an effective energy density and pressure as
\begin{align}
&\rho^\mathrm{(eff)} = T_{00}^\mathrm{(eff)} = \frac{1}{2} \dot{\phi}^2 (1+2\alpha \sqrt{\kappa} \phi ) + (1+4\alpha \sqrt{\kappa} \phi )V\;,\label{eq:rhoeff}\\
&p^\mathrm{(eff)} g_{ij} = T_{ij}^\mathrm{(eff)} = \left[\frac{1}{2} \dot{\phi}^2 (1+2\alpha \sqrt{\kappa} \phi ) - (1+4\alpha \sqrt{\kappa}\phi )V  \right] g_{ij}\;.\label{eq:peff}
\end{align}
Substituting the FRW metric and Eqs.(\ref{eq:rhoeff}) and (\ref{eq:peff}) in the modified Einstein equation (\ref{eq:EinsteinMG}), we can recast the Friedmann equations into the same form as Eqs.(\ref{eq:FriedmannGR}) and (\ref{eq:accelerationGR}) in terms of $\rho^\mathrm{(eff)},~p^\mathrm{(eff)}$ as follows:
\begin{align}
& H^{2}=\frac{\kappa \rho^\mathrm{(eff)}}{3}=\frac{\kappa}{3}\left[\frac{\dot{\phi}^{2}}{2} (1+2\alpha \sqrt{\kappa} \phi )+(1+4\alpha \sqrt{\kappa} \phi )V\right] \;, \\
& \frac{\ddot{a}}{a}=-\frac{\kappa}{6}\left(3 p^\mathrm{(eff)}+\rho^\mathrm{(eff)}\right)=-\frac{\kappa}{3}\left[\dot{\phi}^{2} (1+2\alpha \sqrt{\kappa} \phi )-(1+4\alpha \sqrt{\kappa} \phi )V\right] \;.
\end{align}
Notice that the $\phi$ dependent terms appear in the equations. On the other hand, the modified Klein-Gordon equation is
\begin{align}
\left(\ddot{\phi}+3H\dot{\phi}\right)\left(1+2\alpha\sqrt{\kappa}\phi\right)+\alpha\sqrt{\kappa}\dot{\phi}^2+\left(1+4\alpha\sqrt{\kappa}\phi\right)V_{,\phi}+4\alpha\sqrt{\kappa}V=0\,.\label{KGflrw}
\end{align}
Directly applying the slow-roll approximation, one requires
\begin{align}
\dot{\phi}^2 \ll V,\quad\ddot{\phi} \ll H\dot{\phi},\quad \sqrt{\kappa}\dot{\phi}^2 \ll H\dot{\phi} \;.
\end{align}
Then we get the modified Klein-Gordon equation and the Friedman equation in the slow-roll regime
\begin{align}
&3H\dot{\phi}\left(1+2\alpha\sqrt{\kappa}\phi\right)+\left(1+4\alpha\sqrt{\kappa}\phi\right)V_{,\phi}+4\alpha\sqrt{\kappa}V\simeq0\,,\label{eq:KGslr}\\
& H^{2} \simeq \displaystyle\frac{\kappa}{3}(1+4\alpha \sqrt{\kappa} \phi )V \;. \label{eq:Friedmanslr}
\end{align}
For the slow-roll parameters defined via the potential, we find their analytical expressions as follows:
\begin{align}
\epsilon_V &\simeq \displaystyle\frac{1}{2\kappa (1+2\alpha\sqrt{\kappa}\phi)}\left(\frac{V_{,\phi}}{V}+\frac{4\alpha\sqrt{\kappa}}{1+4\alpha\sqrt{\kappa}\phi}\right)^2 \nonumber\\
&= \displaystyle \frac{1}{2\kappa} \left(\frac{V_{,\phi}}{V} \right)^2 +\alpha \frac{1}{\sqrt{\kappa}}\left[ \frac{4V_{,\phi}}{V}-\phi \left(\frac{V_{,\phi}}{V} \right)^2 \right] + \mathcal{O}(\alpha^2)\;, \label{eq:epsilonMG}\\
\eta_V  &\simeq \displaystyle \frac{1}{\kappa (1+2\alpha\sqrt{\kappa}\phi)} \left[\frac{V_{,\phi\phi}}{V}+ \frac{\alpha\sqrt{\kappa} (7+12\alpha\sqrt{\kappa}\phi)}{(1+2\alpha\sqrt{\kappa}\phi) (1+4\alpha\sqrt{\kappa}\phi)}\frac{V_{,\phi}}{V} -\frac{4\alpha^2\kappa}{(1+2\alpha\sqrt{\kappa}\phi) (1+4\alpha\sqrt{\kappa}\phi)}\right]\nonumber\\
&=  \displaystyle \frac{V_{,\phi\phi}}{\kappa V} +\alpha \frac{7V_{,\phi} -2\phi V_{,\phi\phi}}{\sqrt{\kappa} V} +\mathcal{O}(\alpha^2)  \;, \label{eq:etaMG}
\end{align}
where in the last row of each expression, we expand it assuming a small $\alpha$. In the limit $\alpha\rightarrow0$, the slow-roll parameters in Eqs.(\ref{eq:epsilonMG}) and (\ref{eq:etaMG}) return to their GR counterparts in Eq.(\ref{eq:etaGR}) as expected. As is mentioned before and will be shown later, the values of $\alpha$ to accommodate the inflationary model predictions with observations are usually very small, so the expansions to the linear order in $\alpha$ give good approximations in these cases. Therefore, we will present the expansion in $\alpha$ whenever is possible as a reference to the small $\alpha$ limit.
The number of e-folds is 
\begin{align}
N = \displaystyle \int^{\phi_\mathrm{end}}_{\phi_*} \frac{H}{\dot{\phi}} \mathrm{d}\phi &\simeq \int_{\phi_\mathrm{end}}^{\phi_*} \displaystyle \frac{(1+2\alpha\sqrt{\kappa}\phi) (1+4\alpha\sqrt{\kappa}\phi)\kappa V}{(1+4\alpha\sqrt{\kappa}\phi) V_{,\phi}+4\alpha\sqrt{\kappa}V} \mathrm{d}\phi \nonumber\\
&\simeq \displaystyle\int_{\phi_\mathrm{end}}^{\phi_*} \displaystyle \left[ \frac{\kappa V}{V_{,\phi}} + \alpha \frac{2\kappa^{3/2} V (\phi V_{,\phi} -2V)}{V_{,\phi}^2} \right] \mathrm{d}\phi \;. \label{eq:Nmg}
\end{align}
We omit the higher-order terms in $\alpha$ in the last expression. With the field value evaluated at the horizon crossing using Eq.(\ref{eq:Nmg}), one can calculate the observables $n_s$ and $r$ using Eq.(\ref{eq:nsrGR}).

Before closing this section, we would like to mention that the formulation of the theory with a $f(\phi)T$ coupling can be understood from another point of view. More explicitly, one can redefine a new matter Lagrangian $\mathcal{\tilde{L}}_m$ as
\begin{equation}
\mathcal{\tilde{L}}_m\equiv \alpha\sqrt{\kappa}\phi T+\mathcal{L}_m\,.
\end{equation}
One can further define a new scalar field $\tilde\phi$ and its potential $\tilde V$, such that the matter Lagrangian $\mathcal{\tilde{L}}_m$ can be expressed as the canonical form of this new scalar field:
\begin{equation}
\mathcal{\tilde{L}}_m\equiv-\frac{1}{2}g^{\mu\nu}\partial_\mu\tilde\phi\partial_\nu\tilde\phi-\tilde{V}\,.
\end{equation}
From this perspective, the whole theory can be interpreted as Einstein gravity minimally coupled to the scalar field $\tilde\phi$ with a nontrivial potential. In particular, if we follow the standard definition of the slow-roll parameters:
\begin{equation}
\epsilon_V=\frac{1}{2\kappa}\left(\frac{\tilde{V}_{,\tilde\phi}}{\tilde{V}}\right)^2\,,\qquad \eta_V=\frac{1}{\kappa}\frac{\tilde{V}_{,\tilde\phi\tilde\phi}}{\tilde{V}}\,,
\end{equation}
and use the mapping between $(\phi,V)$ and $(\tilde\phi,\tilde{V})$, one can exactly obtain Eqs.~\eqref{eq:epsilonMG} and \eqref{eq:etaMG} given above. This provides us with a new viewpoint to understand the theory, although the motivation of this nontrivial potential $\tilde{V}$ may be less physically sound.

\section{Application to inflationary models}\label{sec:application}
\subsection{Chaotic inflation}

We consider first the chaotic inflation model~\cite{Linde:1983rmu}, where the potential is of the power-law form
\begin{align}
V=\lambda M_\mathrm{pl}^4 \left( \displaystyle \frac{\phi}{M_\mathrm{pl}} \right)^n \;,
\end{align}
where $n$ is the power index and $\lambda$ is a dimensionless coupling constant. For $\lambda \ll 1$, inflation is insensitive to the initial condition, thus dubbed ``chaotic". It is one of the simplest forms of single-field potentials, and a representative of large field models.  Its possible realization in supergravity has been discussed in~\cite{Kawasaki:2000yn}. This kind of models feature a relatively large tensor-to-scalar ratio, and $n\geq2$ is strongly disfavored by Planck 2018~\cite{Planck:2018jri}.

\begin{figure}[t!]
\centering
\includegraphics[width=.6\textwidth]{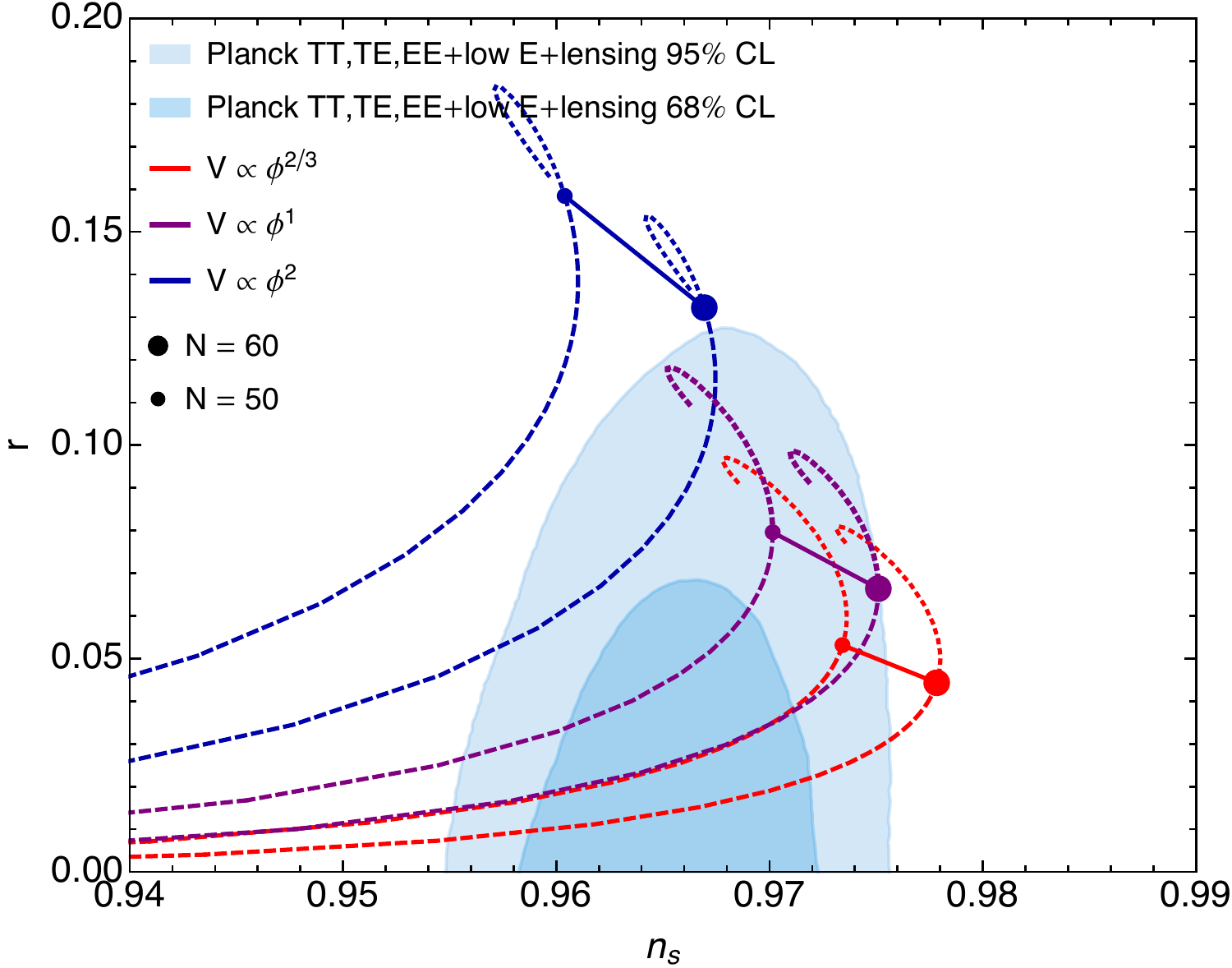}
\caption{The spectral tilt and tensor-to-scalar ratio predicted by the chaotic inflation models in the modified gravity. 
The Einstein gravity results (solid lines) are shown for comparison. The range of $\alpha$ is $[-0.014,1]$. The dashed curves are results in the modified gravity with a negative $\alpha$ while the dotted curves correspond to a positive $\alpha$.  Marginalized joint $68\%$ and $95\%$ C.L. regions for $n_s$ and $r$ at $k = 0.002~ \mathrm{Mpc}^{-1}$ from Planck 2018 data~\cite{Planck:2018jri} are shown in blue and light blue, respectively.}
\label{fig:model_PL}
\end{figure}

In Einstein gravity, with the potential we can write the slow-roll parameters in Eq.~(\ref{eq:etaGR}) as
\begin{align}
\epsilon_\mathrm{E} = \frac{1}{2\kappa} \frac{n^2}{\phi^2} \;, \quad
\eta_\mathrm{E} = \frac{1}{\kappa} \frac{n(n-1)}{\phi^2} \;.
\end{align}
In the $f(\phi)T$ modified gravity, substituting the potential into Eq.~(\ref{eq:epsilonMG}) and Eq.~(\ref{eq:etaMG}), we get the slow-roll parameters
\begin{align}
\epsilon_V&=\displaystyle \frac{1}{2 \kappa  \left(1+2 \alpha  \sqrt{\kappa } \phi \right)} \left(\frac{n}{\phi }+\frac{4 \alpha  \sqrt{\kappa }}{1+4 \alpha  \sqrt{\kappa } \phi }\right)^2\nonumber\\
&= \displaystyle \frac{n^2}{2 \kappa  \phi ^2}+\alpha \frac{  \left(4 n-n^2\right)}{\sqrt{\kappa } \phi }+\mathcal{O}(\alpha ^2) \;,\\
\eta_V&=\displaystyle  \frac{1}{\kappa  \left(1+2 \alpha  \sqrt{\kappa } \phi \right)} \left[\frac{(n-1) n}{\phi ^2} +\frac{n \alpha  \sqrt{\kappa } \left(7+12 \alpha  \sqrt{\kappa } \phi \right)}{\phi  \left(1+2 \alpha  \sqrt{\kappa } \phi \right) \left(1+4 \alpha  \sqrt{\kappa } \phi\right)} -\frac{4 \alpha ^2 \kappa }{\left(1+2 \alpha  \sqrt{\kappa } \phi \right) \left(1+4 \alpha  \sqrt{\kappa } \phi \right)}\right] \nonumber\\
&= \displaystyle  \frac{(n-1) n}{\kappa  \phi ^2}+\alpha  \frac{ \left(9 n-2 n^2\right)}{\sqrt{\kappa } \phi } +\mathcal{O}(\alpha ^2) \;.
\end{align}
According to the expansion in $\alpha$, we see explicitly that the expressions of both slow-roll parameters return to the Einstein gravity ones when $\alpha \rightarrow 0$. On the other hand, with a nonzero value of $\alpha$, the slow-roll parameter that violates the slow-roll condition first needs to be determined with care. As it is $\alpha$-dependent, one can no longer get an expression of the field value at the end of inflation, nor manage to express the slow-roll parameters in terms of the e-folds, i.e., $N$.
The spectral index and tensor-to-scalar ratio can be expressed in terms of $\alpha$ and inflaton field $\phi$ as
\begin{align}
n_s =&~ \displaystyle 1- \frac{n^2}{\kappa \phi ^2}\frac{1}{(1+2 \alpha \sqrt{\kappa} \phi )} 
- \left[\frac{n}{\kappa \phi^2} +4\alpha^2 (4+ 5n)\right] \frac{2}{(1+2 \alpha \sqrt{\kappa}  \phi )^2 (1+4 \alpha \sqrt{\kappa}  \phi )} \nonumber\\
&- \left[\frac{11n}{\sqrt{\kappa}\phi} + 2\alpha (6+22n)\right]  \frac{2\alpha}{(1+2 \alpha \sqrt{\kappa}  \phi )^2 (1+4 \alpha \sqrt{\kappa}  \phi )^2} \nonumber\\
=&~\displaystyle 1- \frac{n^2+2 n}{\kappa  \phi ^2}+ \frac{2 \alpha  \left(n^2-3 n\right)}{\sqrt{\kappa } \phi }+\mathcal{O}(\alpha ^2) \;,\\
r =&~ \displaystyle \frac{8 }{\kappa(1+2 \alpha \sqrt{\kappa}\phi)} \left(\frac{n}{\phi } + \frac{4 \alpha \sqrt{\kappa}}{1+4 \alpha \sqrt{\kappa} \phi }\right)^2 \nonumber\\
=&~\displaystyle \frac{8 n^2}{\kappa  \phi ^2}-\frac{16 \alpha  \left(n^2-4 n\right)}{\sqrt{\kappa } \phi }+\mathcal{O}(\alpha ^2)  \;.\label{eq:chaoticr}
\end{align}
Being conservative, we choose to work with a small $|\alpha| \leq 1$ using the exact expressions and plot the results in Fig.~\ref{fig:model_PL}.

We first focus on $n\leq2$ cases to make a direct comparison with the canonical cases. For $n\leq2$, the end of inflation is determined by $\epsilon_V$ when $\alpha > 0$ as in the case of Einstein gravity, while the slow-roll region is bounded by $\eta_V$ to be in a range of $\alpha$ when $\alpha < 0$. In Fig.~\ref{fig:model_PL}, we see that when $\alpha \rightarrow 0$, the modified gravity predictions agree with that of Einstein gravity. 
A negative $\alpha$ suppresses the tensor-to-scalar ratio (as can be seen from the expansion in Eq.(\ref{eq:chaoticr})) and makes the spectral tilt more red-tilted, and, thus, brings the results into better agreement with observations for certain ranges of $\alpha$. To be more specific, $\alpha$ in $[-0.0105,-0.0079]$ allows the $n=2/3$ line to enter the $1\sigma$ range for $N=60$, and $\alpha$ in $[-0.0094,-0.0059]$ for $N=50$. For $n=1$ case, the values of $\alpha$ in $[-0.0099,-0.0074]$ and $[-0.0089,-0.0040]$ work for $N=60$ and $N=50$ respectively.
Even the strongly disfavored $n=2$ case has some overlap with the Planck $2\sigma$ region with a negative $\alpha$ in the range $[-0.0088,-0.0016]$. 
The $n_s$-$r$ curves are very sensitive to $\alpha$ values. A percent level shift of negative $\alpha$ causes dramatic changes in the $n_s$-$r$ curves. To be compatible with observations, $\alpha$ needs to be carefully chosen within a subpercent range. In short, negative values of $\alpha$ suppresses the tensor-to-scalar ratio in chaotic inflationary models and makes them more consistent with observational data.

Additionally, $n=4$ power-law potential may be more interesting from the viewpoint of particle physics considering a no-scale renormalizable scalar potential. Looking from Fig.~\ref{fig:model_PL}, it seems that there could be a possibility to bring the $n=4$ case into agreement with observations using a negative enough value of $\alpha$. We check that it is not possible as $\eta_V$ constrains the range of $\alpha$ to be in a narrow range. Only in that range we can have slow-roll, but the modification is not large enough to  bring the $n_s$-$r$ lines into the Planck $2\sigma$ region. We can also look at this in another way: $n=4$ leads to a vanishing linear $\alpha$ term in the expansion in $\alpha$ in Eq.(\ref{eq:chaoticr}), and, thus, diminishes the effect of $\alpha$ in lowering $r$, as the leading order in this case is at least $\alpha^2$.

\subsection{Natural inflation}

Natural inflation~\cite{Freese:1990rb,Adams:1992bn} is a well-motivated inflation model as the flatness of the inflationary potential required by the slow-roll is guaranteed by a quasishift symmetry. In natural inflation, inflaton is an axionlike particle and has a potential of the following form:
\begin{align}
V=\Lambda^4 \left[1+\cos \left(\frac{\phi }{f}\right)\right]\;,
\end{align}
where $\Lambda$ is the inflationary energy scale, and $f$ is the decay constant.

\begin{figure}[t!]
\centering
\includegraphics[width=.6\textwidth]{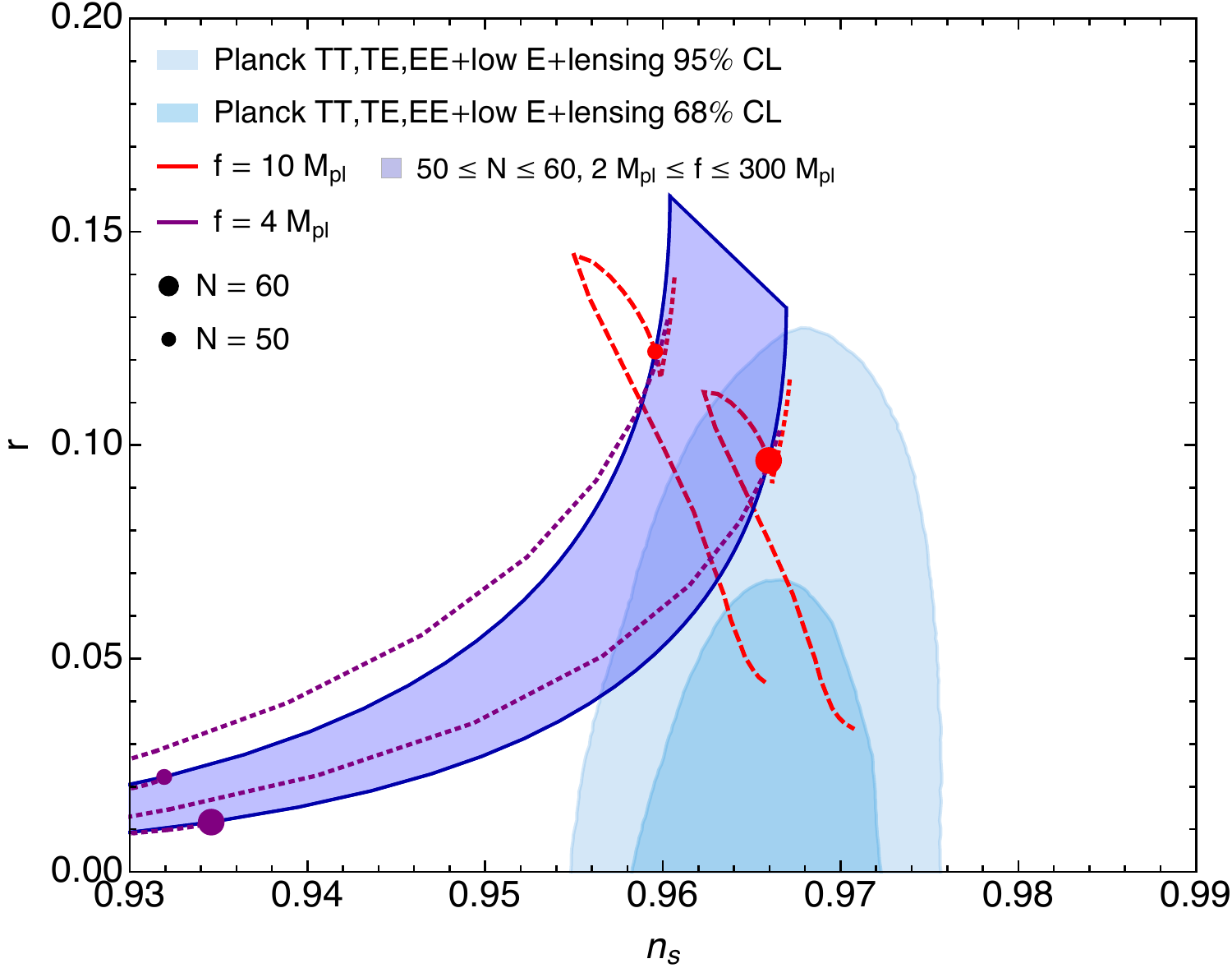}
\caption{ The spectral tilt and tensor-to-scalar ratio predicted by natural inflation models in the modified gravity. The Einstein gravity results (dark blue band) are shown for comparison. The range of $\alpha$ is $[-0.05,0.1]$ for $f=10 M_\mathrm{pl}$ and  $[0,1]$ for $f=4 M_\mathrm{pl}$. 
The dashed lines are results in the modified gravity with a negative $\alpha$ while the dotted lines are plotted with a positive $\alpha$. Marginalized joint $68\%$ and $95\%$ C.L. regions for $n_s$ and $r$ at $k = 0.002~ \mathrm{Mpc}^{-1}$ from Planck 2018 data~\cite{Planck:2018jri} are shown in blue and light blue, respectively.}
\label{fig:model_NI}
\end{figure}

In Einstein gravity, the slow-roll parameters are 
\begin{align}
\epsilon_\mathrm{E} = \frac{1}{2\kappa f^2} \left[ \frac{\sin \left(\frac{\phi }{f}\right)}{1+\cos \left(\frac{\phi }{f}\right)} \right]^2\;,\quad 
\eta_\mathrm{E} = -\frac{1}{\kappa f^2} \frac{\cos \left(\frac{\phi }{f}\right)}{1+\cos \left(\frac{\phi }{f}\right)}\;. 
\end{align}
Within the $f(\phi)T$ gravity, the slow-roll parameters are
\begin{align}
\epsilon_V =&~ \frac{1}{2 \kappa  \left(1+2 \alpha  \sqrt{\kappa } \phi \right)} \left[-\frac{\sin \left(\frac{\phi }{f}\right)}{f \left(\cos \left(\frac{\phi }{f}\right)+1\right)} + \frac{4 \alpha  \sqrt{\kappa }}{1+4 \alpha  \sqrt{\kappa } \phi }\right]^2\nonumber\\
=& \frac{\sin ^2\left(\frac{\phi }{f}\right)}{2 f^2 \kappa  \left[\cos \left(\frac{\phi }{f}\right)+1\right]^2}+\frac{\alpha  \left[-\phi  \sin ^2\left(\frac{\phi }{f}\right)-4 f \sin \left(\frac{\phi }{f}\right)-4 f \sin \left(\frac{\phi }{f}\right) \cos \left(\frac{\phi }{f}\right)\right]}{f^2 \sqrt{\kappa } \left[\cos \left(\frac{\phi }{f}\right)+1\right]^2}+\mathcal{O}(\alpha ^2) \;,\\
\eta_V =& -\frac{1}{\kappa  \left(1+2 \alpha  \sqrt{\kappa } \phi \right)} 
\left[\frac{\cos \left(\frac{\phi }{f}\right)}{f^2 \left(\cos \left(\frac{\phi }{f}\right)+1\right)}
+\frac{\sin \left(\frac{\phi }{f}\right)}{f \left(\cos \left(\frac{\phi }{f}\right)+1\right)} 
\frac{\alpha  \sqrt{\kappa } \left(7+12 \alpha  \sqrt{\kappa } \phi \right)}{\left(1+2 \alpha  \sqrt{\kappa } \phi \right) \left(1+4 \alpha  \sqrt{\kappa } \phi \right) } \right. \nonumber\\
~&\left.+\frac{4 \alpha ^2 \kappa }{\left(1+2 \alpha  \sqrt{\kappa } \phi \right) \left(1+4 \alpha  \sqrt{\kappa } \phi \right)} \right] \nonumber\\
=& -\frac{\cos \left(\frac{\phi }{f}\right)}{f^2 \kappa  \left[\cos \left(\frac{\phi }{f}\right)+1\right]}+\frac{\alpha  \left[2 \phi  \cos \left(\frac{\phi }{f}\right)-7 f \sin \left(\frac{\phi }{f}\right)\right]}{f^2 \sqrt{\kappa } \left[\cos \left(\frac{\phi }{f}\right)+1\right]}+\mathcal{O}(\alpha ^2) \;.
\end{align}
We see again that when $\alpha\rightarrow 0$, the results of $f(\phi)T$ modified gravity go to the Einstein gravity ones. The spectral index and tensor-to-scalar ratio in the $f(\phi)T$ gravity can be expressed accordingly
\begin{align}
n_s =&~ \displaystyle1-\frac{\cos \left(\frac{\phi }{f}\right) \sec ^2\left(\frac{\phi }{2 f}\right)}{\kappa  \left(1+2 \alpha  \sqrt{\kappa } \phi \right)f^2} 
-\frac{3 }{\kappa  \left(1+2 \alpha  \sqrt{\kappa } \phi \right) } \left[\frac{\tan \left(\frac{\phi }{2 f}\right)}{f}-\frac{4 \alpha  \sqrt{\kappa }}{1+4 \alpha  \sqrt{\kappa } \phi }\right]^2\nonumber\\
&-\frac{2 \alpha  \left(7+12 \alpha  \sqrt{\kappa } \phi \right) \tan \left(\frac{\phi }{2 f}\right)}{\sqrt{\kappa } \left(1+2 \alpha  \sqrt{\kappa } \phi \right)^2 \left(1+4 \alpha   \sqrt{\kappa } \phi \right)f} 
-\frac{8 \alpha ^2}{\left(1+2 \alpha  \sqrt{\kappa } \phi \right)^2 \left(1+4 \alpha  \sqrt{\kappa } \phi \right)} \nonumber \\
=&~1-\frac{1}{f^2\kappa} \left[ 3 \tan ^2\left(\frac{\phi }{2 f}\right)-\sec ^2\left(\frac{\phi }{2 f}\right)+2 \right] \nonumber\\
&~+\frac{2\alpha}{f^2\sqrt{\kappa}} \left[ 3 \phi  \tan ^2\left(\frac{\phi }{2 f}\right)+5 f \tan \left(\frac{\phi }{2 f}\right)-\phi  \sec ^2\left(\frac{\phi }{2 f}\right)+2 \phi \right] +\mathcal{O}(\alpha ^2)\;,\\
r =&~ \displaystyle \frac{8}{\kappa  \left(1+2 \alpha  \sqrt{\kappa } \phi \right) } \left[\frac{\tan \left(\frac{\phi }{2 f}\right)}{f}-\frac{4 \alpha  \sqrt{\kappa }}{1+4 \alpha  \sqrt{\kappa } \phi }\right]^2 \nonumber\\
=&~\frac{8}{f^2\kappa}\tan ^2\left(\frac{\phi }{2 f}\right) +\frac{16\alpha}{f^2\sqrt{\kappa}} \left[ \phi  \tan ^2\left(\frac{\phi }{2 f}\right)+4 f \tan \left(\frac{\phi }{2 f}\right) \right]+\mathcal{O}(\alpha ^2)\;.
\end{align}

We show the predictions of the spectral tilt and tensor-to-scalar ratio of natural inflation models within the $f(\phi)T$ modified gravity in Fig.~\ref{fig:model_NI} together with the Einstein gravity results. In the canonical case, natural inflation is strongly disfavored. 
Although it was argued that relaxing assumptions on neutrino properties in the analysis of Planck 2015 data leads to a lower tensor-to-scalar ratio such that natural inflation is in a better agreement with data~\cite{Gerbino:2016sgw}, the improvement is lost once the the baryon acoustic oscillation measurements are included.
We show the changes of predictions in modified gravity starting with two benchmark points: $f=10 M_\mathrm{pl}$ and $f=4 M_\mathrm{pl}$. 
A positive $\alpha$ would first slightly suppress the tensor-to-scalar ratio then enhance it when $\alpha$ keeps increasing (see the dotted curves in Fig.~\ref{fig:model_NI}). On the other hand, a negative $\alpha$ increases $r$ first and then suppresses $r$ (dashed curves). In $f=10 M_\mathrm{pl}$ case, a negative $\alpha$ in $[-0.05,-0.0165]$ brings the curve into the Planck $1\sigma$ region for $N=60$, and $\alpha$ in $[-0.05,-0.0245]$ does the same for $50$ e-folds. In $f=4 M_\mathrm{pl}$ case, a positive $\alpha>0.3251$ can also save the model by bringing the curve into the Planck $2\sigma$ region.

\subsection{Starobinsky inflation}

In this subsection, we will consider the effective potential of the Starobinsky inflation~\cite{Starobinsky:1980te} in Einstein frame as
\begin{align}
V=\Lambda^4 \left(1- e^{-\sqrt{\frac{2}{3}} \sqrt{\kappa} \phi} \right)^2 \;.\label{staropo}
\end{align}
The original idea of Starobinsky inflation was realized by including a quadratic term of the Ricci scalar into the Einstein-Hilbert action. Due to the addition of this quadratic term, the theory contains an additional scalar degree of freedom. After a conformal transformation, one can transform the system from its Jordan frame to the Einstein frame. In the Einstein frame, the theory can be interpreted as Einstein gravity minimally coupled to a canonical scalar field with a potential given by Eq.~\eqref{staropo}. Throughout this subsection, we will strictly stay in the Einstein frame, in which one can interpret that the inflaton is minimally coupled to the Einstein-Hilbert action, although the theory is originated from quadratic curvature modifications in the Jordan frame. Therefore, when we include the $f(\phi) T$ corrections into the model later on, the corrections are directly implanted within the Einstein frame, and the theory in the Jordan frame is no longer the original $R^2$ model anymore.

\begin{figure}[t!]
\centering
\includegraphics[width=.6\textwidth]{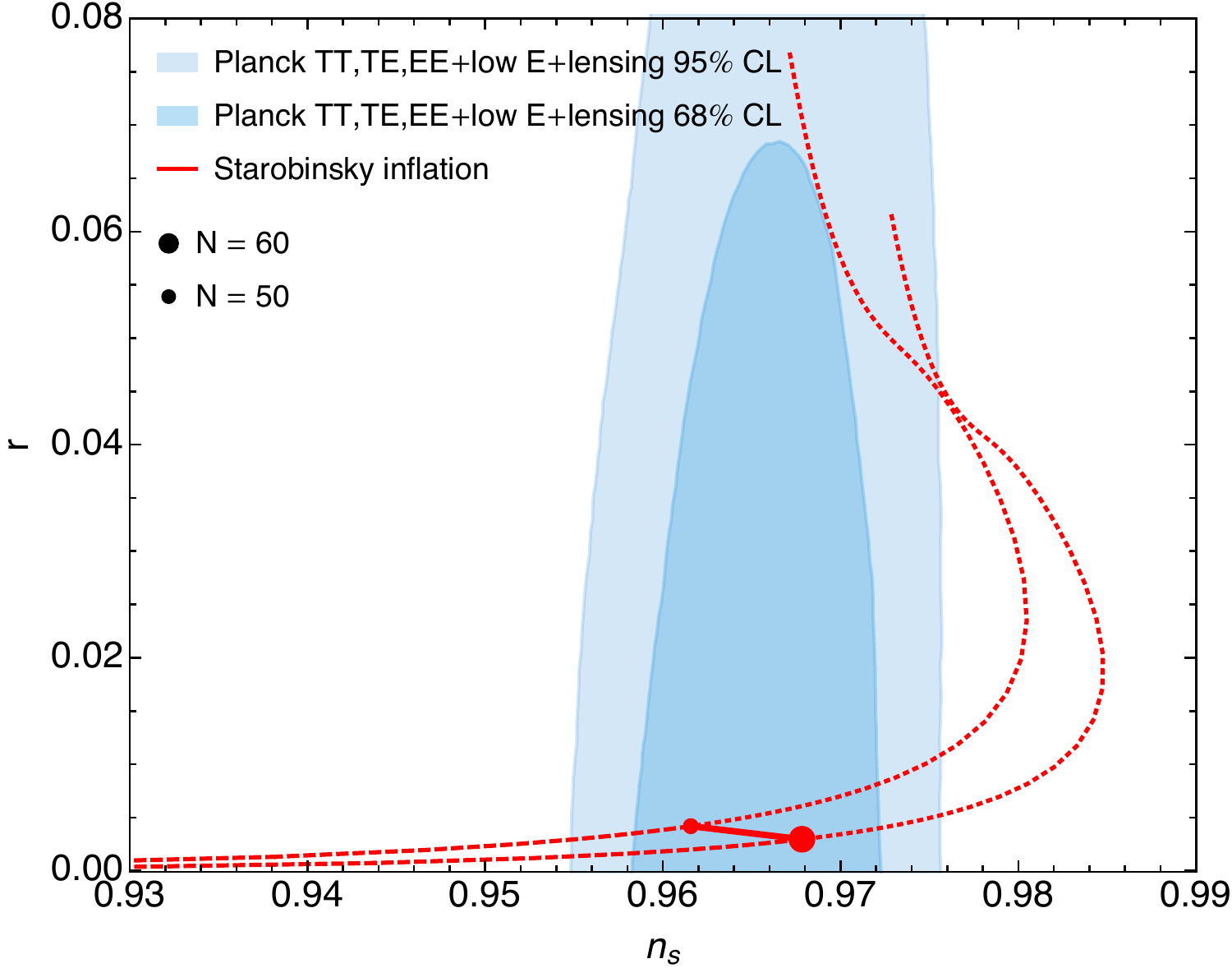}
\caption{The spectral tilt and tensor-to-scalar ratio predicted by the Starobinsky inflation model with $f(\phi)T$ corrections. The results without $f(\phi)T$ corrections are shown for comparison. The range of $\alpha$ is $[-0.01,10]$. The dashed lines are results with a negative $\alpha$ while the dotted lines are plotted with a positive $\alpha$. Marginalized joint $68\%$ and $95\%$ C.L. regions for $n_s$ and $r$ at $k = 0.002~ \mathrm{Mpc}^{-1}$ from Planck 2018 data~\cite{Planck:2018jri} are shown in blue and light blue, respectively.}
\label{fig:model_SI}
\end{figure}

The slow-roll parameters of the Starobinsky model in the Einstein frame read
\begin{align}
\epsilon_\mathrm{E} =~ \displaystyle \frac{4}{3} \left(\frac{e^{-\sqrt{\frac{2}{3}} \sqrt{\kappa}\phi}}{1-e^{-\sqrt{\frac{2}{3}}\sqrt{\kappa}\phi}} \right)^2\;,\quad
\eta_\mathrm{E} =~\displaystyle \frac{4}{3} \frac{\left(2 e^{-\sqrt{\frac{2}{3}} \sqrt{\kappa}\phi}-1 \right)e^{-\sqrt{\frac{2}{3}} \sqrt{\kappa}\phi}}{\left(1-e^{-\sqrt{\frac{2}{3}}\sqrt{\kappa}\phi}\right)^2} \;.
\end{align}
After including the $f(\phi)T$ modification in the Einstein frame, the slow-roll parameters are 
\begin{align}
\epsilon_V =&~\displaystyle \frac{1}{2 \kappa  \left(1+2 \alpha  \sqrt{\kappa } \phi \right)} \left(\frac{2 \sqrt{\frac{2}{3}} \sqrt{\kappa}e^{-\sqrt{\frac{2}{3}} \sqrt{\kappa}\phi }}{1-e^{-\sqrt{\frac{2}{3}} \sqrt{\kappa}\phi }}
+\frac{4 \alpha  \sqrt{\kappa }}{1+4 \alpha  \sqrt{\kappa } \phi }\right)^2 \nonumber\\
=&~\displaystyle\frac{4}{3 \left(e^{\sqrt{\frac{2}{3}} \sqrt{\kappa } \phi }-1\right)^2}+\frac{8 \alpha  \left(-\sqrt{\kappa } \phi +\sqrt{6} e^{\sqrt{\frac{2}{3}} \sqrt{\kappa } \phi }-\sqrt{6}\right)}{3 \left(e^{\sqrt{\frac{2}{3}} \sqrt{\kappa } \phi }-1\right)^2}+\mathcal{O}(\alpha ^2)\;,\\
\eta_V =&~ \displaystyle\frac{1}{\kappa  \left(1+2 \alpha  \sqrt{\kappa } \phi \right)} 
\left[ -\frac{4 \kappa  \left(e^{\sqrt{\frac{2}{3}} \sqrt{\kappa } \phi }-2\right)}{3 \left(e^{\sqrt{\frac{2}{3}} \sqrt{\kappa } \phi }-1\right)^2}
-\frac{4 \alpha ^2 \kappa }{\left(1+2 \alpha  \sqrt{\kappa } \phi \right) \left(1+4 \alpha  \sqrt{\kappa } \phi \right)}\right.\nonumber \\
&~\displaystyle\left. +\frac{\alpha  \sqrt{\kappa } \left(7+12 \alpha  \sqrt{\kappa } \phi \right)}{\left(1+2 \alpha  \sqrt{\kappa } \phi \right) \left(1+4 \alpha  \sqrt{\kappa } \phi \right)} \frac{2 \sqrt{\frac{2}{3}} \sqrt{\kappa } e^{-\sqrt{\frac{2}{3}} \sqrt{\kappa } \phi }}{1-e^{-\sqrt{\frac{2}{3}} \sqrt{\kappa } \phi }}\right]\nonumber\\
=&~\displaystyle-\frac{4 \left(e^{\sqrt{\frac{2}{3}} \sqrt{\kappa } \phi }-2\right)}{3 \left(e^{\sqrt{\frac{2}{3}} \sqrt{\kappa } \phi }-1\right)^2}+\frac{2 \alpha  \left(-8 \sqrt{\kappa } \phi +e^{\sqrt{\frac{2}{3}} \sqrt{\kappa } \phi } \left(4 \sqrt{\kappa } \phi +7 \sqrt{6}\right)-7 \sqrt{6}\right)}{3 \left(e^{\sqrt{\frac{2}{3}} \sqrt{\kappa } \phi }-1\right)^2}+\mathcal{O}(\alpha ^2)\;.
\end{align}
The spectral index and tensor-to-scalar ratio are modified accordingly to be 
\begin{align}
n_s =&~\displaystyle 1-\frac{8 \left(e^{\sqrt{\frac{2}{3}} \sqrt{\kappa } \phi }-2\right)}{3\left(e^{\sqrt{\frac{2}{3}} \sqrt{\kappa } \phi }-1\right)^2 \left(1+2 \alpha  \sqrt{\kappa } \phi \right)}
-\frac{4 \left[\sqrt{6}+\alpha  \left(4 \sqrt{6} \sqrt{\kappa } \phi +6 e^{\sqrt{\frac{2}{3}} \sqrt{\kappa } \phi }-6\right)\right]^2}{3\left(e^{\sqrt{\frac{2}{3}} \sqrt{\kappa } \phi }-1\right)^2 \left(1+2 \alpha  \sqrt{\kappa } \phi \right) \left(1+4 \alpha  \sqrt{\kappa } \phi \right)^2}\nonumber\\
&~\displaystyle+\frac{4 \sqrt{6} \alpha  \left(7+12 \alpha  \sqrt{\kappa } \phi \right)}{3\left(e^{\sqrt{\frac{2}{3}} \sqrt{\kappa } \phi }-1\right) \left(1+2 \alpha  \sqrt{\kappa } \phi \right)^2 \left(1+4 \alpha  \sqrt{\kappa } \phi \right)}
- \frac{24 \alpha ^2}{\left(1+2 \alpha  \sqrt{\kappa } \phi \right)^2 \left(1+4 \alpha  \sqrt{\kappa } \phi \right)} \nonumber\\
=&~\displaystyle\frac{-14 e^{\sqrt{\frac{2}{3}} \sqrt{\kappa } \phi }+3 e^{2 \sqrt{\frac{2}{3}} \sqrt{\kappa } \phi }-5}{3 \left(e^{\sqrt{\frac{2}{3}} \sqrt{\kappa } \phi }-1\right)^2}+\frac{4 \alpha  \left(4 \sqrt{\kappa } \phi +e^{\sqrt{\frac{2}{3}} \sqrt{\kappa } \phi } \left(4 \sqrt{\kappa } \phi -5 \sqrt{6}\right)+5 \sqrt{6}\right)}{3 \left(e^{\sqrt{\frac{2}{3}} \sqrt{\kappa } \phi }-1\right)^2}+\mathcal{O}(\alpha ^2)\;,\\
r =&~\displaystyle\frac{32}{9 \left(e^{\sqrt{\frac{2}{3}} \sqrt{\kappa } \phi }-1\right)^2 \left(1+2 \alpha  \sqrt{\kappa } \phi \right) \left(1+4 \alpha  \sqrt{\kappa } \phi \right)^2}  \left[\sqrt{6}+\alpha  \left(4 \sqrt{6} \sqrt{\kappa } \phi +6 e^{\sqrt{\frac{2}{3}} \sqrt{\kappa } \phi }-6\right)\right]^2 \nonumber\\
=&~\displaystyle\frac{64}{3 \left(e^{\sqrt{\frac{2}{3}} \sqrt{\kappa } \phi }-1\right)^2}+\frac{128 \alpha  \left(-\sqrt{\kappa } \phi +\sqrt{6} e^{\sqrt{\frac{2}{3}} \sqrt{\kappa } \phi }-\sqrt{6}\right)}{3 \left(e^{\sqrt{\frac{2}{3}} \sqrt{\kappa } \phi }-1\right)^2}+\mathcal{O}(\alpha ^2) \;.
\end{align}

We present predictions of the spectral tilt and tensor-to-scalar ratio in Fig~\ref{fig:model_SI}. The results without $f(\phi)T$ corrections, i.e., the original Starobinsky results, are also shown for comparison.
Though the original Starobinsky inflation fits well with data, the $f(\phi)T$ correction with a positive $\alpha$ leads to an enhanced tensor-to-scalar ratio, allowing increased testability of the model. A negative $\alpha$ shifts the lines to the left, i.e., a smaller spectral tilt and tensor-to-scalar ratio. Considering only the lines in the Planck-allowed $2\sigma$ region, we find that $\alpha$ is constrained in the ranges $[-0.0026,0.0031] \cup [6.7694,10]$ when $N=60$ and $[-0.0012,0.0076] \cup [0.2582,10]$ when $N=50$, respectively. 

Comparing the current results with previous preferred ranges of $\alpha$ that bring the $n_s$-$r$ curves into the $1\sigma$ Planck region in chaotic or natural inflation models, we find that different ranges of $\alpha$ are preferred for different models. Chaotic inflation models need a subpercent level of negative $\alpha$, while natural inflation models need a percent level of negative $\alpha$ in the large-field range. For the natural inflationary models, tens of percent level of positive $\alpha$ can even bring the prediction in the small-field range into $2\sigma$ contour of Planck. Finally, Starobinsky model with a subpercent level of positive $\alpha$ can double the value of $r$. With much larger values of $\alpha$, the tensor-to-scalar ratio $r$ can even be enhanced by twenty times with the $f(\phi)T$ modification.

Several other attempts exist to bring inflationary models like chaotic and natural inflation into agreements with Planck results. A warm dissipative effect is considered for natural inflation in Ref.~\cite{Reyimuaji:2020bkm}; a nonminimal coupling to the Ricci scalar also works for natural inflation~\cite{Reyimuaji:2020goi}; UV complete quadratic gravity~\cite{Salvio:2019wcp,Salvio:2020axm} is applied to both chaotic and natural inflation; 
combining natural inflation with Starobinsky inflation also leads to a good agreement with observation~\cite{Salvio:2021lka}; 
$f(R,T)$ gravity is discussed for all the three models~\cite{Gamonal:2020itt}. It is worth noticing that the simple form of $f(R,T)$ considered in Ref.~\cite{Gamonal:2020itt} cannot modify the $n_s$-$r$ prediction for both the chaotic and the natural inflation.

\section{Concluding remarks}\label{sec:conclude}
We propose a simple extension of Einstein gravity named $f(\phi) T$ gravity by including a direct coupling of inflaton with the trace of the energy-momentum tensor in the Einstein-Hilbert action. It is inspired from the simplest $f(R,T)$ form and is natural in that after inflation, the model will resume Einstein gravity. We investigate the inflationary dynamics in this $f(\phi) T$ gravity for simplest form of $f(\phi)=\sqrt{\kappa}\phi$. We present the expressions for the slow-roll parameters as well as spectral tilt and tensor-to-scalar ratio on a general basis and then investigate the effects of the $\phi T$ coupling in three inflationary models. We find that for all the three models under consideration, there are regions of the parameter space that are compatible with the Planck constraints at a $2\sigma$ level. Roughly speaking, a negative $\alpha$ suppresses the tensor-to-scalar ratio while a positive $\alpha$ enhances it. The needed values of the coefficient $|\alpha|$ to bring the $n_s$-$r$ lines into Planck $2\sigma$ region is small, meaning these two observables are sensitive to this $f(\phi) T$ term. For chaotic and natural inflation, a better agreement with the observation is achieved with the help of this modified gravity. For the Starobinsky model, a larger tensor-to-scalar ratio is attainable with positive values of $\alpha$.

We perform the analysis on the simplest form of $f(\phi)=\sqrt{\kappa}\phi$, a more general form of $f(\phi)$ will be carried out elsewhere. This modified gravity is devised with an emphasis on early Universe dynamics. By identifying $\phi$ as inflaton, we do not expect its consequences on the cosmological expansion history after inflation. There will be possible modifications to late-time evolution when relaxing this assumption, and we leave it for a future study. In addition, in this work, we have only focused on the scenario where the energy-momentum tensor is solely given by the inflaton. However, other matter fields, such as massive monopoles \cite{Linde:1994hy,Otalora:2018bso} or energetic particles \cite{Bamba:2014wda,BeltranJimenez:2015xxv}, could also be excited at such high energy scales during inflation. These matter fields, in general, have a nonzero $T$ and could contribute to inflationary dynamics through coupling to inflaton in our model. The possibility of having these matter couplings would lead to interesting inflationary phenomenology and will also be considered in our future works.

\section*{Acknowledgement} 
XYZ is supported in part by the National Natural Science Foundation of China under Grants No. $11775232$ and No. $11835013$, and by the CAS Center for Excellence in Particle Physics.
CYC is
supported by the Institute of Physics of Academia Sinica.


\providecommand{\href}[2]{#2}\begingroup\raggedright\endgroup

\end{document}